\documentclass{article}
\usepackage[T1]{fontenc} 
\usepackage[utf8]{inputenc} 
\usepackage{ismir,amsmath,cite,url,amssymb}
\usepackage{booktabs}
\usepackage{mathtools}
\usepackage{graphicx}
\usepackage[bookmarks=false]{hyperref}
\usepackage{color}
\usepackage{threeparttable}
\usepackage{tabularx}
\usepackage{siunitx}
\newcommand{\norm}[1]{\left\lVert#1\right\rVert}
\usepackage{etoolbox}
\newrobustcmd\ubold{\DeclareFontSeriesDefault[rm]{bf}{b}\bfseries}
\robustify\itshape

\usepackage{lineno}

\title{Neural Waveshaping Synthesis}

\oneauthor
  {Ben Hayes, Charalampos Saitis, George Fazekas}
  {Centre for Digital Music, Queen Mary University of London\\{\tt \{b.j.hayes, c.saitis, g.fazekas\}@qmul.ac.uk}}

\begin{document}

\maketitle
\begin{abstract}
We present the Neural Waveshaping Unit (NEWT): a novel, lightweight, fully causal approach to neural audio synthesis which operates directly in the waveform domain, with an accompanying optimisation (FastNEWT) for efficient CPU inference.
The NEWT uses time-distributed multilayer perceptrons with periodic activations to implicitly learn nonlinear transfer functions that encode the characteristics of a target timbre.
Once trained, a NEWT can produce complex timbral evolutions by simple affine transformations of its input and output signals.
We paired the NEWT with a differentiable noise synthesiser and reverb and found it capable of generating realistic musical instrument performances with only 260k total model parameters, conditioned on F0 and loudness features.
We compared our method to state-of-the-art benchmarks with a multi-stimulus listening test and the Fréchet Audio Distance and found it performed competitively across the tested timbral domains.
Our method significantly outperformed the benchmarks in terms of generation speed, and achieved real-time performance on a consumer CPU, both with and without FastNEWT, suggesting it is a viable basis for future creative sound design tools.
\end{abstract}
\vspace{-6pt}
\section{Introduction}\label{sec:introduction}
\vspace{-2pt}

Synthesisers are indispensable tools in modern music creation.
Over the last six decades, their evolving sonic affordances have defined uncountable musical aesthetics and cultures, enabling composers, sound designers, and musicians to interact with human auditory perception in previously impossible ways.

The recent proliferation of deep neural networks as audio synthesisers is further expanding the capabilities of these tools: realistic instrument performances can be synthesised from simple, low dimensional control signals \cite{engel_ddsp_2020, kim_neural_2019, michelashvili_hierarchical_2020}; the timbre of one instrument can be convincingly transferred to another \cite{engel_ddsp_2020, michelashvili_hierarchical_2020, huang_timbretron_2019, jain_att_2020}; instruments can be morphed and interpolated along nonlinear manifolds \cite{engel_neural_2017, esling_bridging_2018}; and sounds can be manipulated using high level descriptors of perceptual characteristics \cite{esling_bridging_2018, esling_flow_2020, nistal_drumgan_2020}.
Yet despite their impressive abilities, these systems have not been widely adopted in music creation workflows.

We argue that this is largely a pragmatic issue.
Modern music production centres around the digital audio workstation (DAW), with software instruments and signal processors represented as real-time plugins.
These allow users to dynamically manipulate and audition sounds, responsively tweaking parameters as they listen or record.
Neural audio synthesisers do not currently integrate elegantly with this environment, as they rely on deep neural networks with millions of parameters, and are often incapable of functioning in real-time on a CPU.

In this work we move towards integrating the benefits of neural audio synthesis into creative workflows with a novel, lightweight architecture built on the principles of digital waveshaping synthesis \cite{le_brun_digital_1979}.
Our model implicity learns a bank of continuous differentiable waveshapers, which are applied to an exciter signal.
A control module learns to generate time-varying timbres by dynamically shifting and scaling the learnt waveshaper's input and output.
As the waveshapers encode information about the target timbre, our model can synthesise convincing audio using an order of magnitude fewer parameters than the current state-of-the-art methods.

This paper is laid out as follows. In section \ref{sec:related} we discuss related work on neural audio synthesis and waveshaping. Section \ref{sec:nws} introduces our architecture, and we outline our training methodology in section \ref{sec:method}. In section \ref{sec:evaluation} we present and discuss evaluations of our model in comparison to the current state of the art methods \cite{engel_ddsp_2020, michelashvili_hierarchical_2020}.
Finally, we conclude with suggestions for future work in section \ref{sec:conclusion}.
We provide full source code\footnote{\url{https://github.com/ben-hayes/neural-waveshaping-synthesis}} and encourage readers to listen to the audio examples in the online supplement\footnote{\url{https://ben-hayes.github.io/projects/nws/}}.

\vspace{-6pt}
\section{Related Work}\label{sec:related}

\vspace{-2pt}
\subsection{Neural Audio Synthesis}\label{sec:nas}
\vspace{-2pt}

Audio synthesis with deep neural networks has received considerable attention in recent years.
Autoregressive models such as WaveNet \cite{oord_wavenet_2016} and SampleRNN \cite{mehri_samplernn_2017} defined a class of data-driven, general-purpose vocoder, which was subsequently expanded on with further probabilistic approaches, including flow-based models \cite{prenger_waveglow_2019, song_efficient_2020, oord_parallel_2017} and generative adversarial networks \cite{kumar_melgan_2019, kong_hifi-gan_2020, yamamoto_parallel_2020, donahue_adversarial_2019}.
These models allow realistic synthesis of speech, and applications to musical audio \cite{engel_neural_2017, engel_gansynth_2019, hantrakul_fast_2019} have yielded similarly impressive results. 
A parallel stream of research has focused on  controllable musical audio synthesis \cite{esling_bridging_2018, esling_flow_2020, kim_neural_2019, jonason_control-synthesis_2020, engel_ddsp_2020, michelashvili_hierarchical_2020}, in which models are designed to provide control affordances that may be of practical use.
Such controls have included MIDI scores \cite{kim_neural_2019, jonason_control-synthesis_2020}, semantic or acoustical descriptors of timbre \cite{esling_flow_2020, esling_bridging_2018}, and F0/loudness signals \cite{engel_ddsp_2020, michelashvili_hierarchical_2020}.
The representations of timbre learnt by these models have also been observed to show similarities to human timbre perception \cite{hayes_perceptual_2020}.



A recent category of model, \cite{engel_ddsp_2020, wang_neural_2020, michelashvili_hierarchical_2020} unified under the conceptual umbrella of differentiable digital signal processing (DDSP) \cite{engel_ddsp_2020}, has enabled low-dimensional, interpretable control through strong inductive biases to audio synthesis.
Whereas generalised neural vocoders must learn from scratch to produce the features that typify audio signals, such as periodicity and harmonicity, DDSP methods utilise signal processing components designed to produce signals exhibiting such features. 
These components are expressed as differentiable operations directly in the computation graph, effectively constraining a model's outputs to a subspace defined by the processor's capabilities. 

DDSP methods fall into two groups: those where the network generates control signals for a processor, and those where the network is trained to be a signal processor itself.
The DDSP autoencoder \cite{engel_ddsp_2020} falls into the first category as it generates control signals for a spectral modelling synthesiser \cite{serra_spectral_1990}. 
The neural source-filter (NSF) approach \cite{wang_neural_2020, zhao_transferring_2020, michelashvili_hierarchical_2020} is in the second category. It learns a nonlinear filter that transforms a sinusoidal exciter to a target signal, guided by a control embedding generated by a separate encoder.
In other words: the control module ``plays'' the filter network.

The NSF filter network transforms its input through amplitude distortion, as each activation function acts as a nonlinear waveshaper.
A given layer's ability to generate a target spectrum is thus bounded by the distortion characteristics of its activation function.
For this reason, neural source-filter models are typically very deep: Wang et al.'s simplified architecture \cite{wang_neural_2020} requires 50 dilated convolutional layers, and Michelashvili \& Wolf's musical instrument model \cite{michelashvili_hierarchical_2020} consists of 120 dilated convolutional layers -- 30 for each of its four serial generators.

Our method avoids the need for such depth by learning continuous representations of detailed waveshaping functions as small multilayer perceptrons.
These functions are optimised such that their amplitude distortion characteristics allow them to produce spectral profiles appropriate to the target timbre.
This allows our model to accurately transform an exciter signal considerably more efficiently, whilst still exploiting the benefits of the network-as-synthesiser approach.


\vspace{-4pt}
\subsection{Digital Waveshaping Synthesis}\label{sec:waveshaping}
\vspace{-2pt}

In \textit{waveshaping synthesis} \cite{le_brun_digital_1979}, timbres are generated using the amplitude distortion properties of a nonlinear shaping function $f \colon \mathbb{R} \mapsto \mathbb{R}$, which is memoryless and shift invariant. 
Due to its nonlinearity, $f$ is able to introduce new frequency components to a signal \cite{reiss_overdrive_2015}.
When a pure sinusoid $\cos \omega n$ is used as the input to $f$, only pure harmonics are introduced to the signal.
An exciter signal with multiple frequency components, conversely, would result in intermodulation distortion, generating components at frequencies $a\omega_1 \pm b\omega_2, \; \forall a,b \in \mathbb{Z}^+$, for input frequencies $\omega_1$ and $\omega_2$.
This would result in inharmonic components if $\omega_1$ and $\omega_2$ are not harmonically related.

The shaping function $f$ is designed to produce a specific spectral profile when excited with $\cos \omega n$.
This is achieved as a weighted sum of Chebyshev polynomials of the first kind, which possess the property that the $k$th polynomial $T_k$ directly transforms a sinusoid to its $k$th harmonic: $T_k(\cos \omega n) = \cos \omega k n$.
With a function specified in this way, we can define a simple discrete time waveshaping synthesiser

\begin{equation}
\small
x[n] = N[n]f(a[n] \cos \omega n),
\end{equation}
where $a[n]$ is the distortion index and $N[n]$ is a normalising coefficient.
As the frequency components generated by a nonlinear function vary with input amplitude, varying the distortion index over time allows us to generate evolving timbres, whilst the normalising coefficient allows us to decouple the frequency content and overall amplitude envelope of the signal.



\vspace{-6pt}
\section{Neural Waveshaping Synthesis}\label{sec:nws}
\vspace{-4pt}

\begin{figure*}
    \centering
    \includegraphics[width=0.9\textwidth]{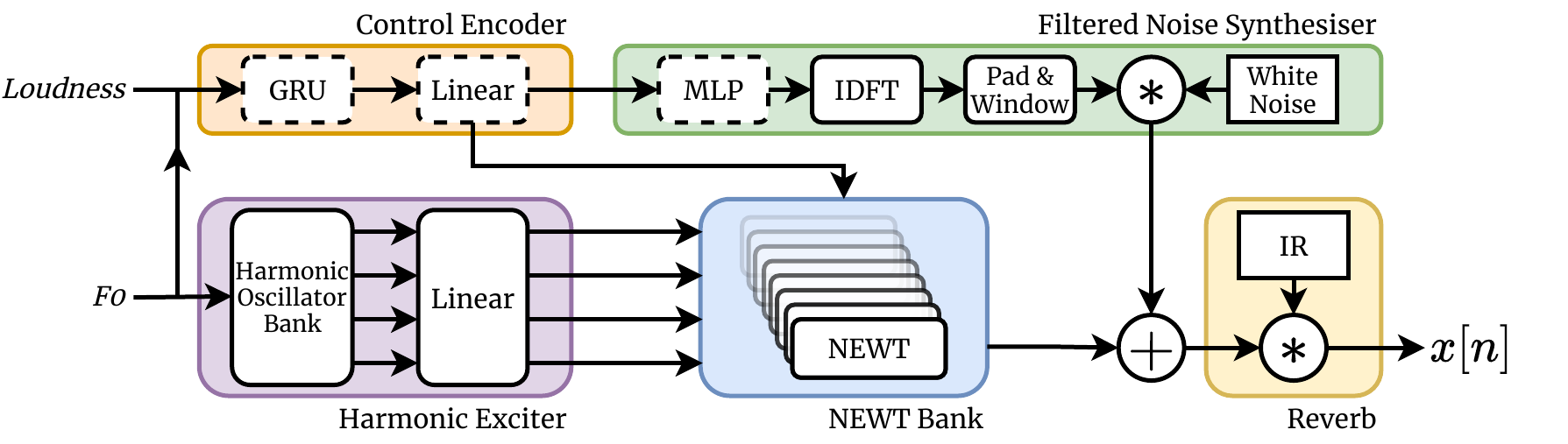}
    \caption{The full architecture of our neural audio synthesiser. All linear layers and MLPs are time distributed. Convolution is denoted $\ast$ and applied by multiplication in the frequency domain. Blocks with dashed outlines operate at the same coarse time steps as the control signal, whilst those with solid outlines operate at audio rate.}
    \label{fig:architecture}
\end{figure*}

Our model acts as a harmonic-plus-noise synthesiser \cite{serra_spectral_1990}.
This architecture separately generates periodic and aperiodic components and exploits an inductive bias towards harmonic signals.
Fig. \ref{fig:architecture} illustrates the overall architecture of our model.

\vspace{-4pt}
\subsection{Control Encoder}
\vspace{-2pt}

We condition our model on framewise control signals extracted from the target audio with a hop size of 128.
We project these to a 128-dimensional control embedding $z$ using a causal gated recurrent unit (GRU) of hidden size 128 followed by a time distributed dense layer of the same size.
We leave the exploration of the performance of alternative sequence models to future work. 

\vspace{-4pt}
\subsection{NEWT: Neural Waveshaping Unit}
\vspace{-2pt}

\begin{figure}[t]
    \centering
    \includegraphics[width=0.9\columnwidth]{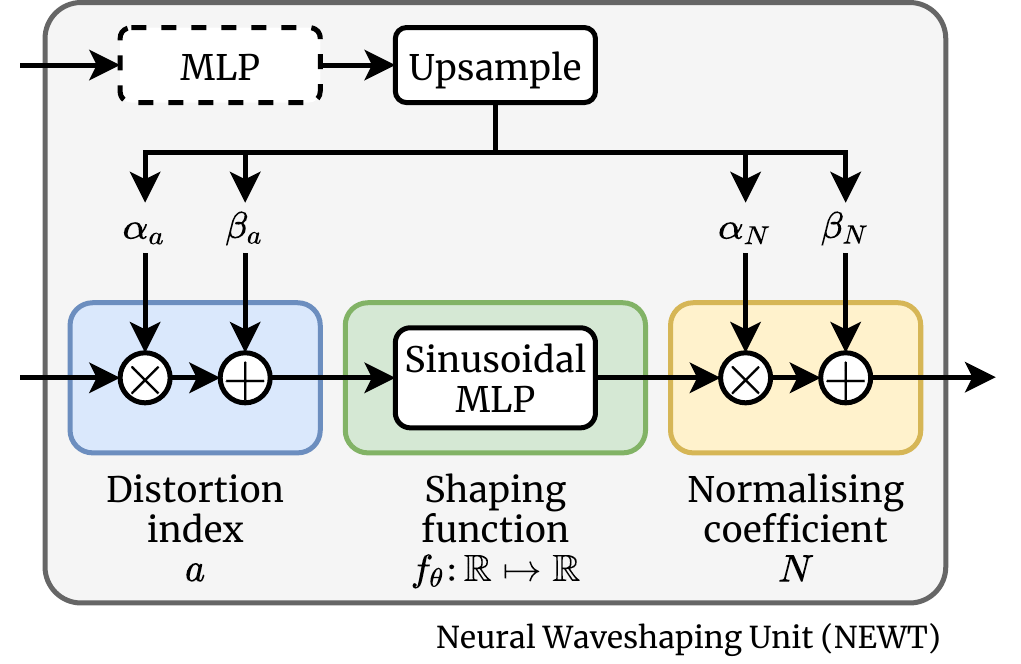}
    \caption{A block diagram depicting the structure of the neural waveshaping unit (NEWT). Blocks with dashed outlines operate at control signal time steps, whilst solid blocks operate at audio rate.}
    \label{fig:newt}
    \vspace{-1\baselineskip}
\end{figure}

The shaping function $f$ of a waveshaping synthesiser can be fit to only a single instantaneous harmonic spectrum.
The spectral evolution afforded by the distortion index $a[n]$ is thus usually unrelated to the target timbre.
This is a limitation of the Chebyshev polynomial method of shaping function design.
Here, we propose to instead learn a shaping function $f_\theta$ parameterised by a multilayer perceptron (MLP).
As demonstrated in recent work on implicit neural representations \cite{sitzmann_implicit_2020, romero_ckconv_2021}, MLPs with sinusoidal activations dramatically outperform ReLU MLPs in learning continuous representations of detailed functions with arbitrary support.
We therefore use sinusoidal activations in $f_\theta$, which enables useful shaping functions to be learnt by very compact networks.
Here, we use 64 parallel shaper MLPs, each with 4 layers, with a hidden size of 8 neurons.

To enable our model to fully exploit the distortion characteristics of $f_\theta$, we replace the distortion index $a[n]$ and normalising coefficient $N[n]$ with affine transforms before and after the shaping function.
The parameters of these transforms, denoted $\alpha_a$ and $\beta_a$ for the distortion index and $\alpha_N$ and $\beta_N$ for the normalising coefficient, are generated by a separate MLP (depth 4, width 128, ReLU activations with layer normalisation \cite{ba_layer_2016}) which takes $z$ as input, and then upsampled to audio rate. 
The output of a single NEWT in response to exciter signal $y[n]$ is thus given by:

\begin{equation}
    \small
    x[n] = \alpha_N f_\theta (\alpha_a y[n] + \beta_a) + \beta_N.
\end{equation}

In this way, the NEWT disentangles two tasks: it learns a synthesiser parameterised by $(\alpha_a, \alpha_N, \beta_a, \beta_N)$, and it learns to ``play'' that synthesiser in response to a control signal $z$.
Fig. \ref{fig:newt} illustrates the structure of the NEWT.
In practice, we use multiple such units in parallel.
We can implement this efficiently using grouped 1-dimensional convolutions with a kernel size of 1 --- essentially a bank of parallel time-distributed dense layers.

\vspace{-4pt}
\subsection{FastNEWT}
\vspace{-2pt}

The NEWT is an efficient approach to generating time-varying timbres, but its reliance on grouped 1-dimensional convolutions best suits it to GPU inference.
Many use-cases for our model do not guarantee the availability of a GPU, and so efficient CPU inference is of crucial importance.
For this reason, we propose an optimisation called the \textit{FastNEWT}: 
as each learnable shaping function simply maps $\mathbb{R} \mapsto \mathbb{R}$, it can be replaced by a lookup table of arbitrary resolution.
Forward passes through $f_\theta$ are then simply replaced with the $\mathcal{O}(1)$ operation of reading values from an array and calculating an interpolation.

To produce a \textit{FastNEWT}, we sample $f_\theta$ across a closed interval.
The sampling resolution and interval are tunable parameters of this operation, and represent a trade-off between memory cost and reconstruction quality.
Here, we opt for a lookup table of 4096 samples over the interval $[-3, 3]$, using a naïve implementation with linear interpolation.
Like the rest of our model, this is implemented using PyTorch operations, and so we treat this as an upper bound on the computational cost of the \textit{FastNEWT}.
In practice, an implementation in a language with low level memory access would confer performance improvements.

\vspace{-4pt}
\subsection{Harmonic Exciter}
\vspace{-2pt}

To reduce the resolution required of the shaping functions, we produce our exciter with a harmonic oscillator bank generating up to 101 harmonics, truncated at the Nyquist frequency.
The outputs of this oscillator bank are passed through a time distributed linear layer, acting as a mixer which provides each NEWT channel with a weighted mixture of harmonics.
Thus, the $i$th output channel of the exciter module is given by:

\begin{equation}
\small
    y_i[n] = \sum_{k=1}^{K} A(k\omega) w_{ik}\cos k \omega n + b_i,
\end{equation}
where the antialiasing mask $A(k\omega)$ is 1 if $-\pi < k\omega < \pi$ and 0 otherwise.

\vspace{-4pt}
\subsection{Noise Synthesiser}
\vspace{-2pt}

In spectral modelling synthesis \cite{serra_spectral_1990}, audio signals are decomposed into a harmonic portion and a residual portion.
The residual portion is typically modelled by filtered noise, with filter coefficients varying over time according to the spectrum of the residual. 
Here, we use an MLP (depth 4, hidden size 128, ReLU activations with layer normalisation) to generate 256-tap FIR filter magnitude responses conditioned on $z$.
We apply a differentiable window-design method like that used in the DDSP model \cite{engel_ddsp_2020} to apply the filters to a white noise signal.
First, we take the inverse DFT of these magnitude responses, then shift them to causal form, and apply a Hann window to the impulse response.
We then apply the filters to a white noise signal by multiplication in the frequency domain.

\vspace{-4pt}
\subsection{Learnable Reverb}
\vspace{-2pt}

To model room acoustics, we apply a differentiable convolutional reverb to the signal.
We use an impulse response $c[n]$ of length 2 seconds, initialised as follows:

\begin{equation}
\small
c[n] 
\begin{cases}
\sim \mathcal{N}(0; \text{1e-6}), & \text{if } n > 1,\\
= 0, & \text{if } n = 0.
\end{cases}
\end{equation}
$c[n]$ is trainable for $n \geq 1$, whilst the $0$th value is fixed at $0$.
The reverberated signal $(c \ast x)[n]$ is computed by multiplication in the frequency domain, and the output of the reverb is summed with the dry signal.

\vspace{-6pt}
\section{Experiments}\label{sec:method}
\vspace{-4pt}

Our model can be trained directly through maximum likelihood estimation with minibatch gradient descent.
Here we detail the training procedure used in our experiments.





\vspace{-6pt}
\subsection{Loss}
\vspace{-4pt}

We trained our model using the multi-resolution STFT loss from \cite{yamamoto_parallel_2020}.
A single scale of the loss is defined as the expectation of the sum of two terms.
The first is the spectral convergence $L_\text{sc}$ (Eqn. \ref{eq:L_sc})  and the second is log magnitude distance $L_\text{m}$ (Eqn. \ref{eq:L_m}), defined as:

\begin{equation}\label{eq:L_sc}
\small
    L_\text{sc}(x, \hat{x}) = \frac{
    \norm{|STFT_m(x)| - |STFT_m(\hat{x})|}_F
    }{
    \norm{|STFT_m(x)|}_F
    }
\end{equation}
and
\begin{equation}\label{eq:L_m}
\small
    L_\text{m}(x, \hat{x}) = \frac{1}{m}
    \norm{\log |STFT_m(x)| - \log |STFT_m(\hat{x})|}_1
\end{equation}
respectively,
where $\norm{\cdot}_F$ is the Frobenius norm, $\norm{\cdot}_1$ is the L1 norm, and $STFT_m$ gives the short-time Fourier transform with analysis window of length $m$ for $ m \in \left\{ 512, 1024, 2048 \right\}$.
We used the implementation of this loss provided in the {\it auraloss} library \cite{steinmetz_auraloss_2020}.



\vspace{-4pt}
\subsection{Data}
\vspace{-2pt}

We collated monophonic audio files from three instruments (violin, trumpet, \& flute) from across the University of Rochester Music Performance (URMP) dataset \cite{li_creating_2019}, and for each instrument applied the following preprocessing.
We normalised amplitude across each instrument subset, made all audio monophonic by retaining the left channel, and resampled to 16kHz.
We extracted F0 and confidence signals using the full CREPE model \cite{kim_crepe_2018} with a hop size of 128 samples.
We extracted A-weighted loudness using the procedure laid out in \cite{hantrakul_fast_2019} using a window of 1024 samples and a hop size of 128 samples.
We divided audio and control signals into 4 second segments, and discarded any segment with a mean pitch confidence $<0.85$.
Finally, control signals were standardised to zero mean and unit variance. 
Each instrument subset was then split into 80\% training, 10\% validation, and 10\% test subsets.

\vspace{-5pt}
\subsection{Training}
\vspace{-3pt}

\begin{table}
    \centering
    \begin{threeparttable}
    \scalebox{0.95}{
        \begin{tabularx}{0.9\columnwidth}{XcX} &
        \begin{tabular}{l|c}
        \toprule\midrule
           \textbf{Model}  & \textbf{Parameters}  \\
           \midrule
            HTP & 5.6M \\
            DDSP-full & 6M \\
            DDSP-tiny & 280k\tnote{*} \\
            \midrule
            \textit{NWS} & \textit{266k} \\
        \midrule\bottomrule
        \end{tabular} &
        \end{tabularx}
        }
        \begin{tablenotes}\footnotesize
        \item[*] The paper reports 240k \cite{engel_ddsp_2020}, but the official implementation contains a model with 280k parameters.
        \end{tablenotes}
    \end{threeparttable}
    \caption{Trainable parameter counts of models under comparison.}
    \label{tab:params}
\end{table}

We trained our models with the Adam optimiser using an initial learning rate of 1e-3.
The learning rate was exponentially decayed every 10k steps by a factor of 0.9.
We clipped gradients to a maximum norm of 2.0.
All models were trained for 120k iterations with a batch size of 8.

\vspace{-6pt}
\section{Evaluation \& Discusson}\label{sec:evaluation}
\vspace{-2pt}


To evaluate the performance of our model across different timbres, we trained a neural waveshaping model for each instrument subset.
We denote these models \textit{NWS}, specifying the instrument where relevant.
After training, we created optimised models with \textit{FastNEWT}, denoted \textit{NWS-FN}, and included these in our experiments also.

\vspace{-4pt}
\subsection{Benchmarks}
\vspace{-2pt}

We evaluated our models in comparison to two state of the art methods: DDSP \cite{engel_ddsp_2020} and Hierarchical Timbre Painting (referred to from here as \textit{HTP}) \cite{michelashvili_hierarchical_2020}.
We trained these on the same data splits as our model, preprocessed in accordance with each benchmark's requirements.

Two DDSP architectures were used as benchmarks: the ``full'' model, originally used to train a violin synthesiser, and the ``tiny'' model described in the paper's appendices.
Both were trained for 30k iterations as recommended in the supplementary materials.
We denote these \textit{DDSP-full} and \textit{DDSP-tiny}, respectively.
HTP comprises four distinct Parallel WaveGAN \cite{yamamoto_parallel_2020} generators operating at increasing timescales.
We trained each for 120k iterations, as recommended in the original paper.  
Table \ref{tab:params} lists the total trainable parameter counts of all models under comparison.


\vspace{-4pt}
\subsection{Fréchet Audio Distance}
\vspace{-2pt}

\begin{table}
    \sisetup{detect-weight,
        detect-all = true,
         mode=text,
         group-minimum-digits = 4}
    \centering
    \scalebox{0.95}{
    \begin{tabular}{l|S S S}
    \toprule\midrule
         & \multicolumn{3}{c}{\textbf{Fréchet Audio Distance}} \\
         \textbf{Model} &  \multicolumn{1}{c}{Flute} & \multicolumn{1}{c}{Trumpet} & \multicolumn{1}{c}{Violin} \\\midrule
        Test Data       & 0.463                      & 0.327                       & 0.096                      \\\midrule
        HTP             & 6.970                      & 14.848                      & 2.529                      \\
        DDSP-full       & 3.091                      & \ubold 1.391                & \ubold 1.062               \\
        DDSP-tiny       & 3.673                      & 5.301                       & \itshape 2.454             \\\midrule
        NWS             & \ubold 2.704               & \itshape 2.158              & 5.101                      \\
        NWS-FN          & \itshape 2.717             & 2.163                       & 5.091                      \\
    \midrule\bottomrule
    \end{tabular}
    }
    \caption{Fréchet Audio Distance scores for all models using background embeddings computed across each instrument's full dataset. Bold type indicates the best performance in a column and italics the second best.}
    \label{tab:fad}
    \vspace{-1\baselineskip}
\end{table}

The Fréchet Audio Distance (FAD) is a metric originally designed for evaluating music enhancement algorithms \cite{kilgour_frechet_2019}, which correlates well with perceptual ratings of audio quality.
It is computed by fitting multivariate Gaussians to embeddings generated by a pretrained VGGish model \cite{hershey_cnn_2017}.
This process is performed for both the set under evaluation, yielding $\mathcal{N}_e(\mu_e, \Sigma_e)$, and a set of ``background'' audio samples which represent desirable audio characteristics, yielding $\mathcal{N}_b(\mu_b, \Sigma_b)$.
The FAD is then given by the Fréchet distance between these distributions:

\begin{equation}
\small
    F(\mathcal{N}_b, \mathcal{N}_e) = 
    \norm{\mu_b - \mu_e}^2 + tr(\Sigma_b + \Sigma_e - 2\sqrt{\Sigma_b \Sigma_e}).
\end{equation}
Thus, a lower FAD score indicates greater similarity to the background samples in terms of the features captured by the VGGish embedding.
Here, we used the FAD to evaluate the overall similarity of our model's output to the target instrument.
We computed our background embedding distribution $\mathcal{N}_b$ from each instrument's full dataset, whilst the evaluation embedding distributions $\mathcal{N}_e$ were computed using audio resynthesised from the corresponding test set.
FAD scores for our model, all benchmarks, and the test datasets themselves are presented in Table \ref{tab:fad}.

In general, the closely matched scores of the NWS and NWS-FN models indicate that, across instruments, the \textit{FastNEWT} optimisation has a minimal effect on this metric of audio quality.
On trumpet and flute, our models consistently outperform HTP and DDSP-tiny, and also outperform DDSP-full on flute.
On violin, conversely, both DDSP models are the best performers, with HTP achieving a similar score to DDSP-tiny.

\vspace{-4pt}
\subsection{Listening Test}
\vspace{-2pt}


\begin{figure*}
    \centering
    \includegraphics[width=\textwidth]{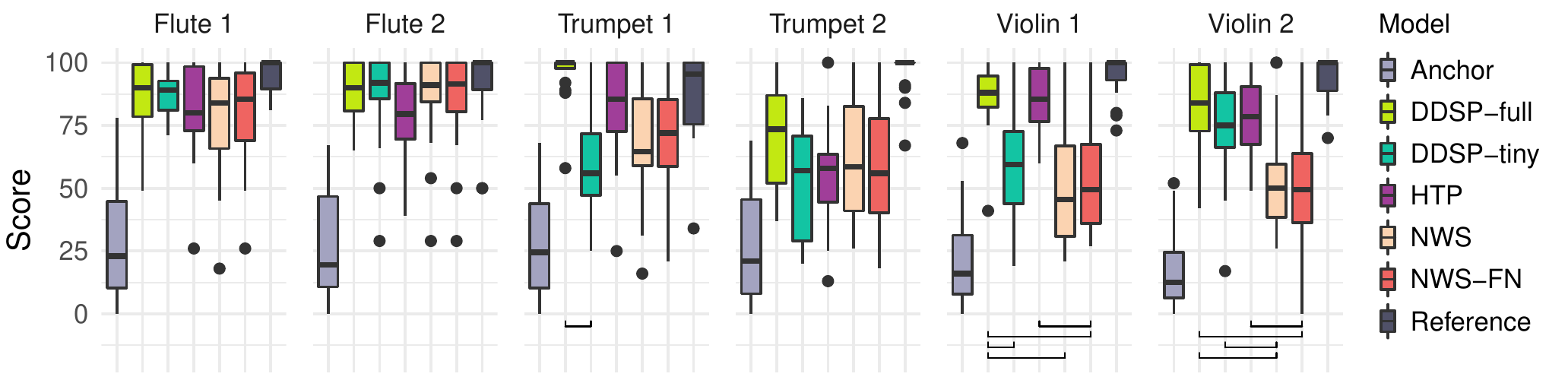}
    \caption{Boxplots of ratings given to each synthesis model during each trial in our listening test. Brackets indicate significant (corrected $p < .005$) differences in pairwise Wilcoxon signed-rank tests with Bonferroni correction.}
    \label{fig:mushra}
    \vspace{-1\baselineskip}
\end{figure*}

Our model and benchmarks can be considered as highly specified audio codecs.
We therefore applied a listening test inspired by the MUSHRA (MUltiple Stimuli with Hidden Reference and Anchor) standard \cite{itu-r_bs1534-3_method_2015}, which is used to assess the perceptual quality of audio codecs.
We used the webMUSHRA framework \cite{schoeffler_webmushra_2018}, adapted to incorporate a headphone screening test \cite{milne_online_2020}.
For each instrument, we selected two stimuli from the test set representing distinct register and articulation, giving six total trials.
In each trial, we used the original recording as the reference and produced the anchor by applying a 1kHz low pass filter.
We recruited 19 participants from a pool of audio researchers, musicians, and audio engineers.
We excluded the responses of one participant, who rated the anchor above the reference in greater than 15\% of trials.
Responses for each trial are plotted in Fig. \ref{fig:mushra}.
In general, NWS and NWS-FN performed similarly across trials, suggesting that FastNEWT has little, if any, impact on the perceptual quality of the synthesised audio.
Across flute and trumpet trials our models were rated similarly to the benchmarks.
In the first violin trial, our models' ratings were similar to those of DDSP-tiny, whilst in the second they were lowest overall.
These ratings are concordant with FAD scores: our model performs competitively on trumpet and flute whilst struggling somewhat with violin.

To examine the influence of melodic stimuli on participants' ratings, we performed Wilcoxon's signed-rank test between scores given for each instrument's two stimuli, for each synthesis model.
For example, scores given to DDSP-full for stimulus Flute 1 were compared to scores given to DDSP-full for Flute 2.
Out of fifteen tests, significant differences ($p < .001$) were observed in two: between trumpet stimuli for both DDSP-full and HTP.
No other significant effects were observed ($\alpha = 0.05$).

To examine the effect of synthesis model, we performed Friedman's rank sum test on ratings from each trial.
For flute stimuli, no significant effects were found.
Significant effects were observed for both trumpet stimuli, although Kendall's $W$ suggested only weak agreement between raters (Trumpet 1: $Q=27.45, p < 0.001, W=0.38$; Trumpet 2: $Q=14.18, p < 0.01, W=0.20$) .
Both violin stimuli also resulted in significant effects with moderate agreement between raters (Violin 1: $Q=42.28, p < 0.001, W=0.59$; Violin 2: $Q=37.95, p < 0.001, W=0.53$).
Post-hoc analysis was performed within each trial using Wilcoxon's signed-rank test with Bonferroni $p$-value correction.
Significant differences (corrected threshold $p < .005$) were observed for Trumpet 1, Violin 1, and Violin 2.
These are illustrated as brackets in Fig. \ref{fig:mushra}.

\vspace{-4pt}
\subsection{Real-time Performance}
\vspace{-2pt}

\begin{table}[t]
    \centering
    \sisetup{detect-weight,
        detect-all = true,
         mode=text,
         group-minimum-digits = 4}

    \scalebox{0.95}{
    \begin{tabular}{l|S[table-format=0.3] S[table-format=0.3]|S[table-format=0.3] S[table-format=0.3]}
    \toprule\midrule
         & \multicolumn{4}{c}{\textbf{Real-time Factor}} \\
         & \multicolumn{2}{c}{GPU} & \multicolumn{2}{c}{CPU} \\
        \textbf{Model} & \multicolumn{1}{c}{\textit{Mean}} & \multicolumn{1}{c}{\textit{90th Pctl.}} & \multicolumn{1}{c}{\textit{Mean}} & \multicolumn{1}{c}{\textit{90th Pctl.}} \\
        \midrule
       HTP & 0.105 & 0.106 & 2.203 & 2.252 \\
       DDSP-full & 0.038 & 0.047 & 0.363 & 0.395 \\
       DDSP-tiny & 0.032 & 0.039 & 0.215 & 0.223 \\
       \midrule
       NWS & \itshape 0.004 & \itshape 0.004  & \itshape 0.194  & \itshape 0.208  \\
       NWS-FN & \ubold 0.003 & \ubold 0.003 & \ubold 0.074 & \ubold 0.076 \\
    \midrule\bottomrule
    \end{tabular}
    }
    \caption{Real-time time factor computed by synthesising four seconds of audio in a single forward pass. Statistics computed over 100 runs. Bold indicates the best performance in a column and italics the second best.}
    \label{tab:rtf_4s}
    \vspace{-4pt}
\end{table}

\begin{figure}[t]
    \centering
    \includegraphics[width=0.9\columnwidth]{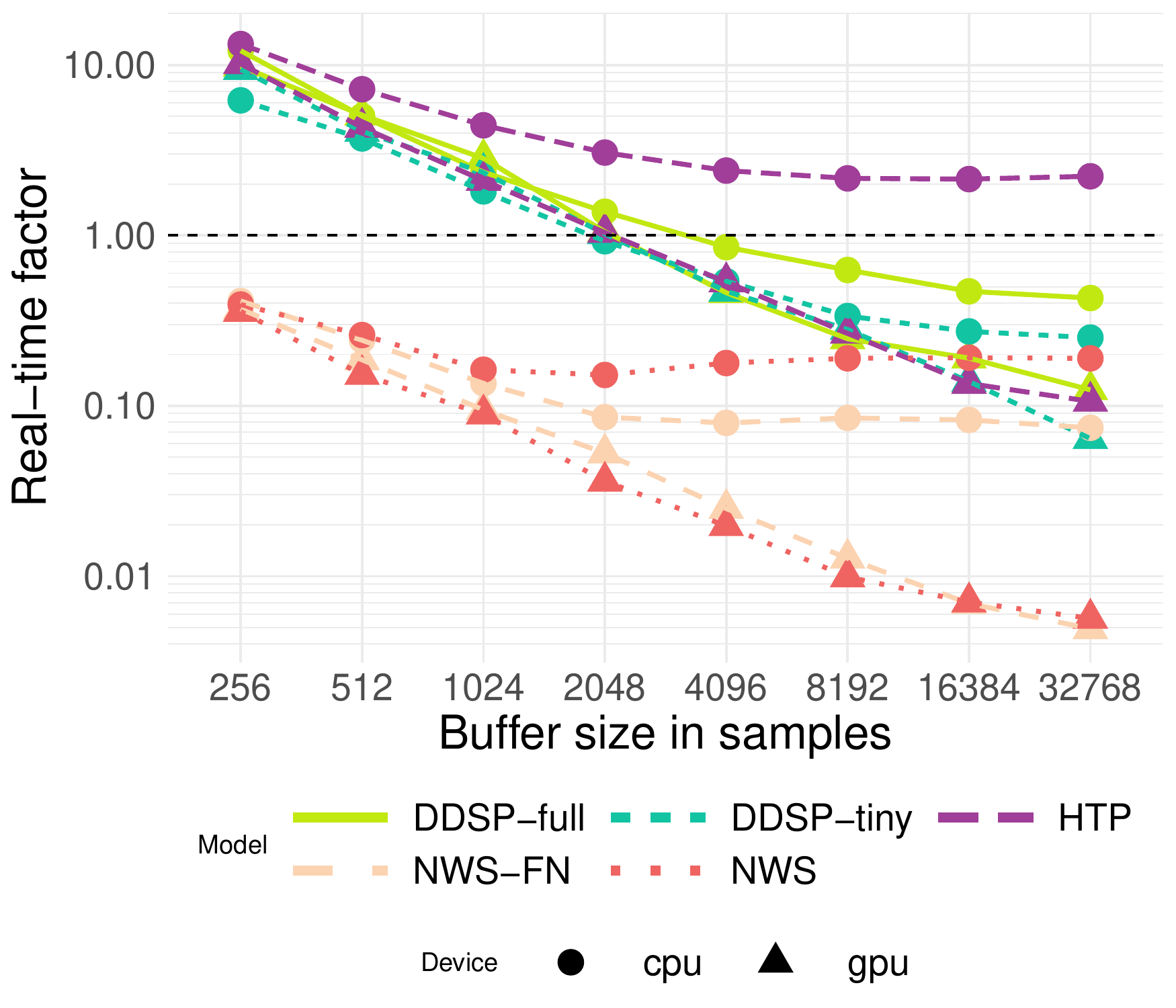}
    \caption{A plot of the mean real-time factor against buffer size across all benchmarks. Mean computed over 100 runs per model per device per buffer size.}
    \label{fig:buffer}
    \vspace{-1\baselineskip}
\end{figure}

We evaluated the real-time performance of our model in two scenarios.
In both cases we took measurements on a GPU (Tesla P100-PCIe 16GB) and a CPU (Intel i5 1038NG7 2.0GHz) and used the real-time factor (RTF) as a metric.
The RTF is defined as

\begin{equation}
\small
    RTF \coloneqq \frac{t_p}{t_i},
\end{equation}
where $t_i$ is the temporal duration of the input and $t_p$ is the time taken to process that input and return an output.
Real-time performance thus requires $RTF<1$.
In all tests we computed RTF statistics over 100 measurements.

The first scenario models applications where an output is expected immediately after streaming an input.
To test this, we computed the RTF on four second inputs.
We report the mean and 90th percentile in Table \ref{tab:rtf_4s}.
On the GPU, NWS and NWS-FN outperformed all benchmarks, including DDSP-tiny.
On the CPU, NWS still outperformed all other models, albeit by a narrower margin.
The benefit of the \textit{FastNEWT} optimisation was clearer on CPU: NWS-FN had a mean RTF $2.9 \times$ lower than the best performing benchmark.
On both platforms, HTP was significantly slower, likely due to its much greater depth.

The second scenario assumes applications where immediate response to input is expected, such as in a software instrument.
Here, samples are processed in blocks to ensure that sufficient audio is delivered to the DAC in time for playback.
We computed the RTF for each buffer size in $B \coloneqq \{ 2^n \mid n \in \mathbb{Z}, 8 \leq n < 16 \}$.
The means of these runs are plotted in Fig. \ref{fig:buffer}.
Again, NWS and NWS-FN outperformed all benchmarks on both CPU and GPU, sitting comfortably below the real-time threshold of 1.0 at all tested buffer sizes.
HTP did not achieve real-time performance at any buffer size on the CPU, and only did so for buffer sizes over 2048 on the GPU.
DDSP-full, similarly, was unable to achieve realtime performance for buffer sizes of 2048 or lower on GPU or CPU, while DDSP-tiny sat on the threshold at this buffer size.
It should be noted that a third-party, stripped down implementation of the DDSP model was recently released, which is capable of real-time inference when the convolutional reverb module is removed\footnote{\url{https://github.com/acids-ircam/ddsp_pytorch}}.




\vspace{-6pt}
\section{Conclusion}\label{sec:conclusion}
\vspace{-2pt}

In this paper, we presented the NEWT: a neural network structure for audio synthesis based on the principles of waveshaping \cite{le_brun_digital_1979}.
We also present full source code, pre-trained checkpoints, and an online supplement containing audio examples.
Our architecture is lightweight, causal, and comfortably achieves real-time performance on both GPU and CPU, with efficiency further improved by the \textit{FastNEWT} optimisation.
It produces convincing audio directly in the waveform domain without the need for hierarchical or adversarial training. 
Our model is also capable of many-to-one timbre transfer by extracting F0 and loudness control signals from the source audio.
Examples of this technique are provided in the online supplement.

In evaluation with a multi-stimulus listening test and the Fréchet audio distance our model performed competitively with state-of-the-art methods with over $20 \times$ more parameters on trumpet and flute timbres, whilst performing similarly to a comparably sized DDSP benchmark on violin timbres.
Due to the use of a harmonic exciter in our architecture and the scope of our experimentation, further work is necessary to ascertain to what degree the NEWT itself contributes to our model's performance.
Therefore, in future work we will perform a full ablation study and a quantitative analysis of the degree to which a trained model makes use of the NEWT's waveshaping capabilities.
In the meantime, the online supplement demonstrates through visualisations of learnt shaping functions, affine parameters $(\alpha_a, \beta_a, \alpha_N, \beta_N)$, and audio taken directly from the output of the NEWT, that the NEWTs in our model do indeed perform waveshaping on the exciter signal.

We suspect the  lower scores on violin timbres were due to the greater proportion of signal energy in higher harmonics in these sounds.
The NEWT may thus been unable to learn shapers capable of producing these harmonics without introducing aliasing artefacts.
Using sinusoidal MLPs with greater capacity inside the NEWT may allow more detailed shaping functions to be learnt, whilst retaining efficient inference with FastNEWT.
Future work will investigate this and other differentiable antialiasing strategies, including adaptive oversampling \cite{de_man_adaptive_2014}.
We will also explore extending our model to multi-timbre synthesis.



\section{Acknowledgements}
We would like to thank the anonymous reviewers at \textit{ISMIR} for their thoughtful comments. We would also like to thank our colleague Cyrus Vahidi for many engaging and insightful discussions on neural audio synthesis. This work was supported by UK Research and Innovation [grant number EP/S022694/1].

\bibliography{references}

%
%
%
%
%

\end{document}